%% file: main.tex
\documentclass[conference,12pt]{IEEEtran}
\usepackage{epsfig,endnotes}
\usepackage{epsfig, endnotes, verbatim, gensymb, textcomp, subcaption, graphics, multirow, amsmath, pslatex, listings, verbatim, amssymb, amsmath, amsthm, alltt, multirow, graphicx, amsfonts, wrapfig, boxedminipage, rotating, color, xspace, setspace, url, lscape, syntax, proof, array, caption, epsfig, endnotes, bbm, mdframed, cite, comment, times, hyperref, cleveref,gensymb, times}

\usepackage[ruled,vlined]{algorithm2e}

\newcommand{\name}{Eyephone\xspace}
\graphicspath{{./Figures/}}

\begin{document}

\title{Secure Mobile Technologies for Proactive\\ Critical Infrastructure Situational Awareness}

\author{
\IEEEauthorblockN{Gabriel Salles-Loustau, Vidyasagar Sadhu, \\ Dario Pompili, Saman Zonouz, Vincent Sritapan$^*$}
\IEEEauthorblockA{
Department of Electrical and Computer Engineering, $^*$Cyber Security Division\\
Rutgers University, $^*$Department of Homeland Security Science \& Technology Directorate\\
\textit{ \{gs643, vidyasagar.sadhu, pompili, saman.zonouz\}@rutgers.edu}, \textit{vincent.sritapan@hq.dhs.gov}
}
}

\IEEEoverridecommandlockouts 
\maketitle

\input{Sources/sec-abstract}
\input{Sources/sec-introduction}

\input{Sources/sec-overview}

\input{Sources/sec-swirls}

\input{Sources/sec-evaluations}

\input{Sources/sec-conclusions}

\footnotesize
\linespread{0.01}
\bibliographystyle{IEEEtran}
\bibliography{references}

\end{document}

%% file: Sources/sec-abstract.tex
\begin{abstract}
Trustworthy operation of our national critical infrastructures, such as the electricity grid, against adversarial parties and accidental failures requires constant and secure monitoring capabilities. In this paper, \name is presented to leverage secure smartphone sensing and data acquisition capabilities and enable pervasive sensing of the national critical infrastructures. %
The reported information by the smartphone users will notify the control center operators about particular accidental or malicious remote critical infrastructure incidents. The reporting will be proactive regarding potentially upcoming failures given the system's current risky situation, e.g., a tree close to fall on a power grid transmission line. The information will include various modalities such as images, video, audio, time and location. \name will use system-wide information flow analysis and policy enforcement to prevent user privacy violations during the incident reportings. %
%
%
A working proof-of-concept prototype of \name is implemented. Our results show that \name allows secure and effective use of smartphones for real-time situational awareness of our national critical infrastructures.
\end{abstract}

%% file: Sources/sec-introduction.tex
\section{Introduction}
\label{sec:introduction}

Critical infrastructures provide essential societal functionalities that are vital for our nation's homeland security and economy~\cite{zonouz2012elimet}. Secure and reliable operation of the critical infrastructures, specifically power grid infrastructures, require effective predictive situational awareness capabilities. Continuous and precise comprehension of the system’s safety status and potential failures will enable operators and/or automated incident response systems to prepare proactively against complex failure scenarios. An example failure would be a cascading blackout launched by a tree fall on a transmission line like what happened in August 2003 North East blackout that affected 50 million Americans and caused approximately 6 billions dollars of damage~\cite{BlackoutReport}. 

To enable real-time monitoring of the critical infrastructures such as the electricity grid, they are instrumented with various types of sensors, e.g., voltage and current sensors so called transformers~\cite{WoodWollenberg}. Such system-level sensors provide the operators with situational awareness regarding \textit{just-occurred} incidents and failures, and hence enable \textit{reactive} responses. However, as shown by the past research, decision-making upon optimal response in large-scale power grid platforms is often time-consuming, e.g., optimal non-linear power flow calculations~\cite{WoodWollenberg}. Please note that the underlying physical power system does not pause and resumes its operation and may enter new states, while the optimization for the best response action is performed for the previous observed state. Consequently, reactive situational awareness often leads to \textit{too-late} notifications. Predictive data acquisition capabilities are needed to forecast upcoming unsafe incidents before they happen. This gives the operators and incident response engines the time they need to proactively come up with optimal countermeasures if those incidents actually happen. 

\begin{figure*}[tp]
  \centering
  \includegraphics[width=\textwidth]{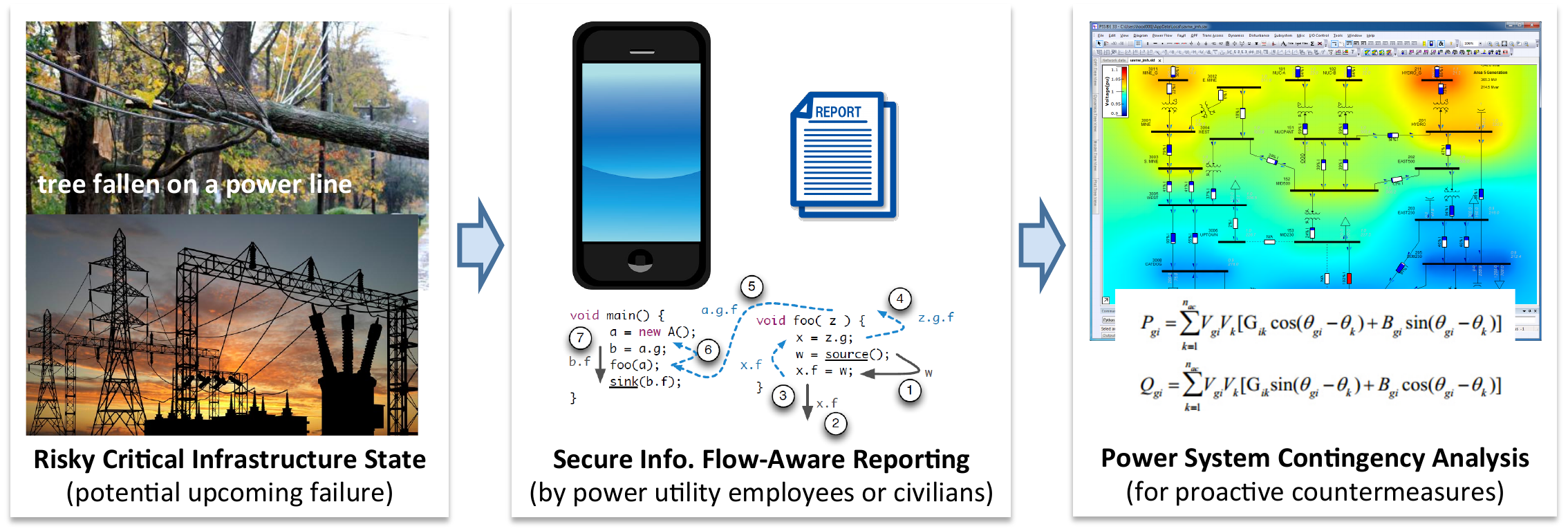}
  \vspace{-0.15in}
  \caption{\name's High-Level Architecture}
  \vspace{-0.15in}
  \label{fig:architecture}
\end{figure*}

There have been several offline risk analysis methods proposed for critical infrastructures that, in theory, could be used for proactive situational awareness. Among the mostly used techniques in practice, power system contingency analysis algorithms~\cite{Ekwue:SelectionReview} attempt to go through \textit{all} possible future failures and perform a ``what-if'' risk analysis. However, as proved practically~\cite{davis:multipleoutage}, enumeration of all possible failures, without prior (probabilistic) knowledge about the incidents, does not scale up for real-world large-scale infrastructures with thousands of power system components such as transmission lines and generators. The reason is that the problem turns into a combinatorial problem with exponential time complexity growth for multiple-failure scenarios, i.e., $\binom{N}{x}$ for considering $x$ failures for an $N$-component power system.  %

In this paper, we present \name, a privacy-preserving crowd-sourcing framework that employs rich smartphone sensing capabilities for distributed reporting of potentially upcoming incidents in power grid infrastructures. \name provides civilians and power utility employees with easy-to-use and user-transparent solutions to use their smartphones to collect and send concise reports about existing risky conditions that may lead to power grid failures in a near future. \name leverages various smartphone sensing modalities for comprehensive real-time reporting. To ensure preservation of the user and employee privacy, \name implements system-wide information flow tracking capabilities, and guarantees no confidential and sensitive data of the user is leaving the phone to the power utility without his/her permission dynamically. \name leverages cryptographic measures to prevent denial-of-service misbehaviors such as malicious spoofed reporting of the fake power grid incidents. %

The contributions of this paper are as follows: 
\begin{itemize}
	\item We propose a smartphone-enabled crowd-sourcing framework to enable predictive monitoring of the critical infrastructures for timely proactive incident response capabilities.
    \item We developed smartphone security and user privacy-preserving solutions, and enhanced power grid contingency analysis algorithm, where the future incidents are selectively analyzed depending on their occurrence probabilities. 
    \item We implemented a working proof-of-concept prototype of our proposed solution in Android operating system, and experimented with it in real-world practical scenarios that gave promising outcomes.
\end{itemize}

%% file: Sources/sec-overview.tex
\section{Proposed Work}
\label{sec:overview}

\autoref{fig:architecture} shows \name's high-level architecture and proposed workflow to facilitate critical infrastructure protection using smartphone-based secure sensing capabilities. The main objective of proactive situational awareness is to detect and report risky situations in critical infrastructures before they cause system failures that later may lead to catastrophic cascading blackouts. For example as shown on the figure, a tree may fall on a transmission line while the power flows through the line safely; however, such a marginal safety may not last long before the line short circuits with the ground and gets detected by the protection power relays at the end of the line. Reporting the tree fall earlier (before it happens) would give the operators more time to assess and decide upon possible corrective actions. Traditional system sensors are not designed to diagnose and report such risky states that may lead to system failures, where the infrastructure operates normally get notified by system-level sensors’ viewpoint. \name provides the power utility operators and civilians with secure smartphone-based solutions and applications to report such risky situations in time for a timely response. 

\name leverages information flow tracking to ensure secure reporting of the risky situations without user's privacy violations. Nowadays, mobile devices are increasingly used in a variety of roles with wildly differing data-protection requirements ranging from the access of sensitive emails and sensor data to the production and sharing of personal content via online social networks. While it is desirable for such diverse content to be accessible from the same device via a unified user experience, today's mobile operating systems provide few, if any, functionalities for fine-grained data protection and isolation. Consequently, heavyweight isolation schemes such as different apps, different virtual machines or, in the extreme case, different devices for accessing different types of content (e.g., risky situation reporting vs. personal) are used, which provide a disharmonious experience for the users. 

\name leverages system-wide dynamic taint analysis to develop a data-protection architecture for mobile operating systems to isolate \emph{data} rather than isolating execution environments and applications. \name tags data based on its security context and controls data mixing between contexts using data flow-based policies. In doing so, it provides users with a unified environment and app developers with an API to construct seamless user interfaces that allow access to data with different security requirements through the same apps without fear of any malicious or inadvertent data leakage.

Secure and reliable maintenance of existing power grid critical infrastructures requires accurate modeling and analysis of the power system dynamics. The three main functionalities that turn a traditional power system into a \textit{smart grid}~\cite{farhangi2010path} are the following: \textit{i)} monitoring that requires instrumentation of the power system components with many sensors to measure various system parameters such as transmission line current magnitudes; \textit{ii)} state estimation and contingency analysis that receive the noisy sensor measurements, calculate noiseless current state of the system using algorithms such as weighted least squares~\cite{abur2004power}, and perform speculative risk analysis of potential upcoming system component failures, so-called \textit{contingencies}. To enable such analyses, critical infrastructure operators and utilities employ power system modeling. The most common type of models used in power system operations are steady state power flow models and sensitivities.  These models are used to monitor the state of the system and predict the effects of changes on the system.

Steady state power system modeling consists of enforcing the conservation of power.  Given a set of power injections and withdrawals, the power flow finds the set of voltages and angles that satisfy the conservation of power.  The system state may be written as
\begin{equation}
\label{eq:powersystemstate}
 \bf{ x = \left[V,\theta\right] }
\end{equation}
where $\bf V$ is a vector of voltages, $\bf \theta$ is a vector of voltage angles.  The vector of real power loads is $\bf P_l$ and the vector of reactive power loads is $\bf Q_l$.  Since generator outputs are controllable (within limits), they are collected separately in a vector of controls, $\bf u$.

The power flow problem can now be written as
\begin{equation}
\label{eq:powerflowsimple}
\bf f(x,u)=0
\end{equation}
where $\bf f(x,u)=0$ is a complex vector representing the injection at each node in the system.  The function $\bf f(x,u)$ represents the system model.  It encapsulates factors like line impedances and system topology. Breaking $\bf f(x,u)$ into real and reactive parts gives
\begin{equation}
\label{eq:powerflowreal}
f_i^p = -P_i^g+P_i^l + \sum_{k \in C}{|V_i||V_k| (G_{ik} \cos{\theta_{ik}} + B_{ik} \sin{\theta_{ik}}})
\end{equation}
\begin{equation}
\label{eq:powerflowreactive}
f_i^q = -Q_i^g+Q_i^l + \sum_{k \in C}{|V_i||V_k| (G_{ik} \sin{\theta_{ik}} - B_{ik} \cos{\theta_{ik}}})
\end{equation}
These equations represent the nonlinear problem that is commonly called the power flow in power systems literature.  The power flow is at the heart of most power systems analysis.  It provides the basis for many tools and sensitives that are used to predict the state of the system in the event of an outage.

%% file: Sources/sec-swirls.tex
\section{Smartphone Data Access Control}
\label{sec:overview}

\begin{figure*}
  \centering
  \includegraphics[width=1\textwidth]{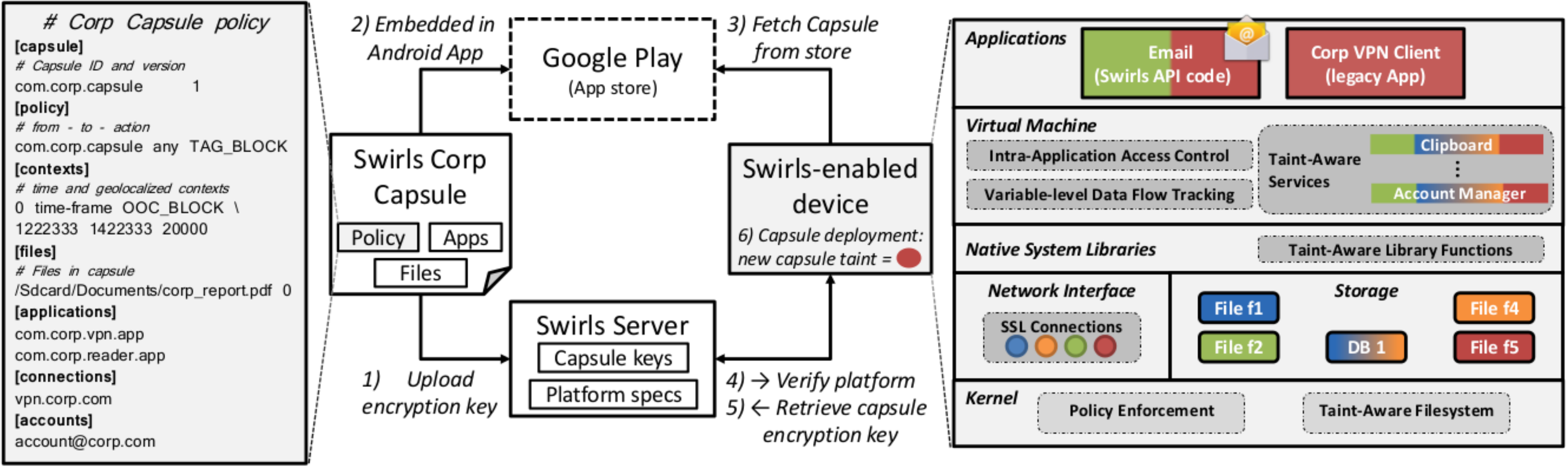}
  \caption{Smartphone Data Access Control}
  \label{fig:overview:design}
\end{figure*}

\autoref{fig:overview:design} shows \name's smartphone-based data flow and access control solution~\cite{zonouz2016swirls}. \name's main objective is to facilitate data separation between the critical infrastructure reports and the user's personal data through deployment of virtual micro security perimeters that we call \textit{capsules}. Each capsule is an encrypted and signed package that includes sensitive data and policies. These policies define how data should be treated when merged with data from other capsules. \autoref{fig:overview:design}'s left block presents a simplified capsule policy. Every capsule is packaged and signed by its corresponding data owner. \name verifies the integrity of the installed capsule signatures on the device before enforcing its associated policies. In the first case, the corporate admins protect their sensitive data, against outsiders, whereas in the second and third cases, the external parties (i.e., developers and third-parties) attempt to protect against curious and potentially careless device users.  

We describe the three major steps in \name. First, for distribution and installation of a capsule, \name implements a secure protocol to prevent malicious capsule modification and/or interception attacks. Second, to protect sensitive data, \name employs dynamic taint analysis to keep track of the installed capsule boundaries while data moves within the system. Third, \name implements efficient mandatory access control at various data propagation points within Android to prevent unauthorized data accesses. 

\noindent\textbf{Capsule distribution and installation.} \name implements a secure capsule distribution and installation interface for nontechnical users and capsule owners. Upon the definition of a capsule, the capsule owner signs and encrypts it. The signatures and encryption keys are pushed to \name's remote server. Capsules can be either downloaded as standalone files or packaged in smartphone app files, and hence are distributable via the app market. The user downloads a signed and encrypted capsule, which is then installed by the \name system app on the smartphone. The installation consists of two steps. First, during a platform verification procedure, \name's remote server verifies the authenticity of the local agent on the system to ensure that capsule policies will be enforced correctly. Second, the \name system app verifies the signature, decrypts the capsule, enforces the capsule policy and installs the capsule data. During the capsule installation, \name allocates a new and unique taint label and dynamically marks the capsule objects as \textit{sources}. The objects can be apps and data files included in the capsule or sensitive data sources such as network connections. 

As a clarification, the \name server does not distribute capsules: it stores decryption keys for capsules. Thus, the phone must authenticate with the server to get the keys in order to decrypt a capsule and access its contents. This design ensures that a phone has \name data protection in place before the content owner makes the content available. The need for OS verification is based on the security requirements of the content owner. Lightweight alternatives such as use of a hardware counter to track flash/root attempts (like Samsung KNOX does) could be used, or checking could be eschewed altogether, as is done for many BYOD apps today.

\noindent\textbf{Capsule boundary tracking.} To guarantee data protection, \name keeps track of capsule boundary growth by tracing the sensitive data propagation starting from the capsule's source objects. \name deploys system-wide taint tracking techniques across various layers of Android to monitor data flow among the following system entities: files, Android content providers, apps, system processes and services, account entries, secure socket connections, interprocess data exchanges, system service calls, incoming network traffic, as well as accounts data. To retain capsule boundary information across smartphone reboots, \name stores references to tainted objects for each capsule in a global database and keeps its information up-to-date whenever new objects are tainted.

\noindent\textbf{Capsule policy enforcement.} The capsule policies mandate how \name should handle access requests to different capsules' data throughout the system. \name's runtime policy enforcement uses the real-time information from the aforementioned capsule boundary database through a three level instrumentation of the Android framework. First, \name controls data accesses within the Linux kernel to ensure that the low-level capsule data propagation complies with the installed policies. This includes filesystem operations and inter-process communications among apps. \name's kernel-level support makes it resilient against malicious access control evasions through Java Native Interface (JNI) code segments. Second, \name instruments the Dalvik virtual machine (VM) layer with policy enforcement modules to control fine-grained access requests to variables within individual apps. \name's kernel-level enforcement is more lightweight than its Dalvik VM counterpart; however, it uses Dalvik layer enforcement for fine-grained access controls when a multi-context app includes data from different capsules simultaneously (\name's kernel-level implementation cannot distinguish different taints within an app).  Finally, \name enhances and controls data interactions among several key system services, e.g., the clipboard service, that are accessed by multiple resources in the system and aggregate data from various sources.

%% file: Sources/sec-evaluations.tex
\section{Performance Evaluation}
\label{sec:evaluations}

We have implemented a real-world implementation of our Android \texttt{4.1.1_6} solution on a test-bed power system environment. We have obtained preliminary results for this extended abstract submission, and will further expand this section for our full paper submission. 

\begin{figure}[tp]
    \centering
    \begin{subfigure}[t]{0.23\textwidth}
        \centering
        \includegraphics[width=\textwidth]{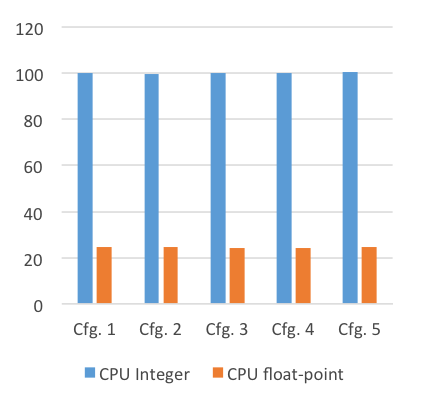}
        \caption{CPU Usage}\label{fig:antutu:proc}
    \end{subfigure}%
    ~ 
    \begin{subfigure}[t]{0.23\textwidth}
        \centering
        \includegraphics[width=\textwidth]{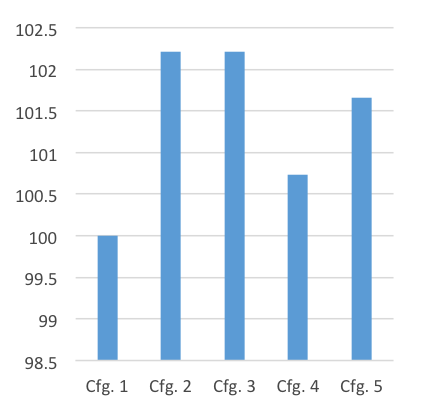}
        \caption{Android Memory}\label{fig:antutu:ram}
    \end{subfigure}
    \caption{Smartphone Resource Utilization}
    \vspace{-0.15in}
\end{figure}

We evaluated the smartphone implementations of \name under five different configurations. We used the Antutu benchmark~\cite{antutu} to quantify \name's overhead on the phone's dynamic computation and memory resource consumptions. The five configurations include: \textit{i)} Cfg. one was the vanilla Android operating system. we also used this configuration as the normalization base for our other results; \textit{ii)} Cfg. two shows the impact of taint analysis and information flow tracking implementations in Android operating system for one of the sensors on the phone's performance; \textit{iii)} Cfg. three is similar to the first configuration except it includes the case where \name creates separate report-specific filesystem objects to ensure separation of the power failure report data, obtained from smartphone sensors such as camera and microphone, from the rest of the phone including the user's sensitive private data; \textit{iv)} Cfg. four shows the following reportings where the filesystem object initialization is already carried out; and \textit{v)} Cfg. five represents the complete run including multiple sensor data tracking dynamically throughout the phone.

\begin{figure*}[tp]
  \centering
  \includegraphics[width=.58\textwidth]{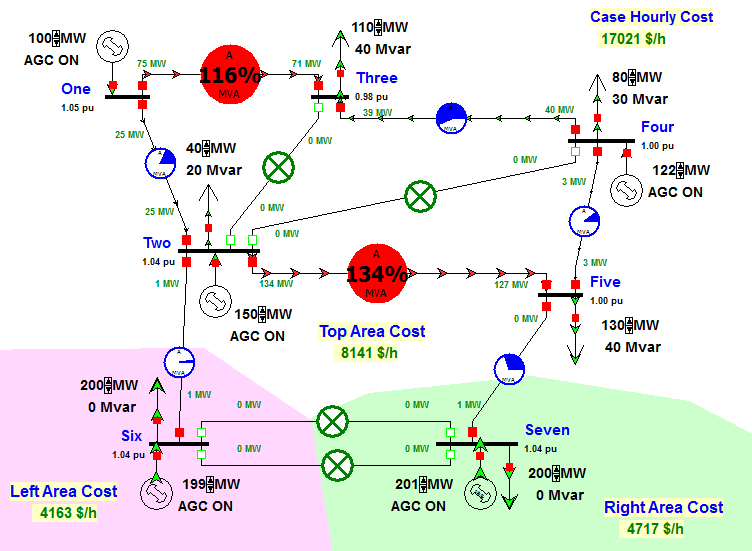}
  \caption{Smartphone-Enabled Contingency Analysis}
  \vspace{-0.15in}
  \label{fig:powersystem}
\end{figure*}

\begin{figure}[tp]
  \centering
  \includegraphics[width=.47\textwidth]{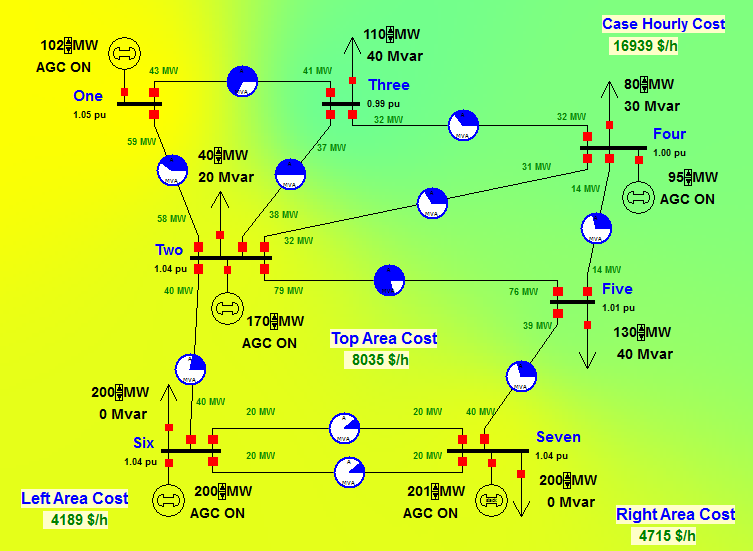}
  \caption{Seven-Bus Power System Test-Bed}
  \vspace{-0.15in}
  \label{fig:powersystem}
\end{figure}

\autoref{fig:powersystem} shows the power system test-bed for our experiments. The power system included seven buses that are spread out and are interconnected through power transmission lines. We modeled the power system using the equations discussed earlier in the paper. \name used the models for predictive contingency analysis given the reports submitted by the smartphone devices. The figure shows the results, i.e., the contour of the system after its contingency analysis. According to the reported risky situations as well as the contingency analysis outcomes, \name detects the south west region of the grid closer to its capacity and more prone to failure, while the east regions are and will remain in safe states. Consequently, the operators need to develop countermeasure strategies proactively before the failure prone regions experience a potential upcoming incident. The development of the countermeasure strategies are outside the scope of this paper.

%% file: Sources/sec-conclusions.tex
\section{Conclusions}
\label{sec:conclusions}

In this extended abstract, we presented \name, a crowd-sourcing framework to enable predictive and scalable situational awareness capabilities for power grid critical infrastructures. \name provides power utility operators and civilians with a secure smartphone solution to report risky situations (possible before they turn into sensor-detectable system failures) to the control rooms for a timely countermeasure in case the failure occurs. Our experimental results showed that \name works effectively on Android $v4.1$. \name protects the user privacy by ensuring no sensitive user data leakage during the reporting through system-wide information flow analysis. 

\section*{Acknowledgments}

We appreciate the Department of Homeland Security Science \& Technology Directorate (DHS S\&T) Cyber Security Division for their support of our project under the contract No. D15PC00159.